\documentclass[11pts]{article}
\usepackage{graphicx}
\usepackage{graphicx}
\usepackage{amsmath}
\usepackage{amssymb}
\usepackage{mathtools}
\usepackage{authblk}
\begin{document}
\title
{Constraining parameters of effective field theory of inflation from Planck data}
\date{ }
\author[1]{Anindya Banerjee}
\author[2]{Himanshu Khanchandani}
\affil[1]{Indian Institute of Technology Kharagpur, Kharagpur, India}
\affil[2]{Indian Institute of Technology Roorkee, Roorkee, India}

\maketitle

\begin{abstract}
The Cosmic Microwave Background can provide information regarding physics of the very early universe, more specifically, of the matter-radiation distribution of the inflationary era. Starting from the effective field theory of inflation, we use the Goldstone action to calculate the three point correlation function for the Goldstone field, whose results can be directly applied to the field describing the curvature perturbations around a de Sitter solution for the inflationary era. We then use the data from the recent Planck mission for the parameters $f_{NL}^{equil}$ and $f_{NL}^{orthog}$ which parametrize the size and shape of non-Gaussianities generated in single field models of inflation. Using these known values, we calculate the parameters relevant to our analysis, $f_{NL}^{\dot{\pi}^3}$, $f_{NL}^{\dot{\pi}(\partial _i \pi)^2}$ and the speed of sound $c_s$ which parametrize the non-Gaussianities arising from two different kinds of generalized interactions of the scalar field in question.
\end{abstract}

\section{Introduction}
\paragraph{ }
The Cosmic Microwave Background(CMB) is an important probe of physics of the early universe. We study the phenomenon of primordial inflation and its observational signatures on the CMB spectrum(for review see \cite{Baumann,Mukhanov,Smoot}). In standard inflationary scenarios, the universe should be very close to a Gaussian random field. The two point correlation function and its Fourier transform, the angular power spectrum for most models of inflation give similar prediction of a scale invariant, adiabatic, Gaussian spectrum. Therefore, in order to distinguish between the competing models of inflation and to constrain the parameters common to these models, we look at the bispectrum predictions from these models, and deviations from Gaussianity. Such a calculation of three-point function was first done by Maldacena in \cite{Maldacena}. The theory of single field slow roll inflation achieves accelerated expansion by means of a scalar field slowly rolling down a potential. It predicts non-Gaussianities that should be too small for observation. However, other models of inflation predict larger departures from the Gaussian spectrum, and have their characteristic types of non-Gaussianities. Any observation of large non-Gaussianity from the Planck mission thus allows us to constrain these models. Moreover, the angular bispectrum has different shapes, and the different models of inflation show peaks for distinct shapes. As elaborated in \cite{regan,Creminelli2} and references therein, various models of primordial non-
Gaussianity are known as local, equilateral, orthogonal or 
folded models in literature. Different aspects of physics of 
the early universe appear in different shapes of the three 
point function.

\begin{itemize}

\item \textbf{Local Non-Gaussianity} appears in multi-field models of 

inflation due to interactions which operate on superhorizon scales.

\item \textbf{Equilateral Non-Gaussianity} includes single field models with non-
canonical kinetic term such as k-inflation or Dirac-Born-
Infeld inflation models \cite{DBItong,DBI2,DBI3,DBI4} characterized by more general higher 
derivative interactions of the inflaton field such as ghost 
inflation and models arising from effective field theories.

\item \textbf{Folded Non-Gaussianity} include single field models with non Bunch Davies vacuum.

\item \textbf{Orthogonal Non-Gaussianity} may be generated in single field models of inflation with a non-canonical kinetic term or with general 
higher derivative interactions. The orthogonal form is 
constructed in such a way that it is nearly orthogonal to both 
local and equilateral forms.\cite{Komatsu}

The plan of this article is as follows: In the next section we review the effective field theory model of inflation. After that we 
explicitly calculate 2-point and 3-point function in section 3 and finally conclude in 
section 4. 

\end{itemize}
\section{The Effective Field Theory of inflation}
\paragraph{}
Cheung, Creminilli, Fitzpatrick, Kaplan and Senatore have used 
the effective field theory approach in \cite{Creminelli} to describe 
the theory of fluctuations around an inflating cosmological 
background. While the inflaton field $\phi$ is a scalar under 
all diffeomorphisms, the perturbation $\delta \phi$ is a scalar only under spatial diffeomorphisms, and transforms non-linearly with respect to time diffeomorphisms,
$$t \rightarrow t+\xi ^0(t,\vec{x})$$
$$\delta \phi \rightarrow \delta \phi + \dot{\phi _0}\xi ^0$$
We can describe the perturbations during  inflation directly around the time-evolving vacuum where the time diffeomorphisms are non-linearly realised. In unitary gauge, the most generic Lagrangian with broken time diffeomorphisms and unbroken spatial diffeomorphisms around a flat FRW with Hubble parameter $H(t)$  is given by \cite{Creminelli},
$$S = \int d^4 x \sqrt{-g} [\frac{1}{2}M_{pl}^2 R + M_{pl}^2 \dot{H}g^{00}- M_{pl}^2(3H^2 + \dot{H}) + \frac{M_2 ^4 (t)}{2!}(g^{00} + 1)^2 $$
$$ + \frac{M_3 ^4 (t)}{3!}(g^{00} + 1)^3+ \ldots  +\frac{M_2 ^2 (t)}{2}( \delta K_{\mu} ^{\mu})^2 + ... ]
$$
\paragraph{}
As elaborated by Senatore, Smith and Zaldariagga, in inflation there is a physical clock that controls the end of inflation, so that time translations are spontaneously broken, and there is a Goldstone boson associated with the symmetry breaking. The Lagrangian of the Goldstone boson is highly constrained by the symmetries of the problem, in this case the fact that spacetime is approximately de Sitter, and $\frac{\dot{H}}{H^2}<<1 $. The Goldstone boson, $\pi$ can be thought of as being equivalent, in standard models of inflation driven by a scalar field, to the perturbations in the scalar field $\delta \phi$. The relation valid at linear order is $\pi = \delta \phi / \dot{\phi}$, where $\dot{\phi}$ is the speed of the background solution. The Goldstone boson is related to the standard curvature perturbation $\zeta$ by the relation, $\zeta =- H\pi$, which is valid at linear order and leading order in the generalised slow roll parameters. The most general Lagrangian for the Goldstone boson is given by \cite{Creminelli,Senatore1,Senatore2} 
The general action for the perturbation $\pi$ is 
\begin{equation} \label{eq: action}
S_{\pi} = \int \mathrm{d^4} x\sqrt{-g}\left[-\frac{M_{pl}^2 \dot{H}}{c_s^2}\left( \dot{\pi}^2-c_s^2 \frac{(\partial_i \pi)^2}{a^2}\right)  + M_{pl}^2 \dot{H}\left(1-\frac{1}{c_s^2}\right) \left( \dot{\pi}^3 - \dot{\pi}\frac{(\partial_i \pi)^2}{a^2}\right) -\frac{4}{3}M_3^4\dot{\pi}^3 + ......\right]   
\end{equation}
\paragraph{}
To arrive at the Lagrangian for the Goldstone mode from the 
most generic Lagrangian in unitary gauge, one performs a time 
diffeomorphism with parameter $\xi ^0 (t,x)$ and promotes the 
parameter to a field, $\pi (t,x)$, which shifts under time 
diffeomorphisms, $\pi (t,\vec{x}) \rightarrow \pi (t.\vec{x}) 
+ \xi ^0 (t,\vec{x})$. This scalar, $\pi$, is the Goldstone 
mode which non-linearly realises time diffeomorphisms and 
describes the scalar perturbations around the FRW solution \cite{Creminelli}.
\section{Calculation of Non-Gaussianities}
We use the action described above in Eq. \ref{eq: action} here to calculate the bispectrum.

\paragraph{Solution of quadratic action}
Considering only upto the quadratic terms the action is,
\begin{equation*}
S_2 = \int \mathrm{d^4} x(-M_{pl}^2) \dot{H}\left(\frac{a^3 \dot{\pi}^2}{c_s^2}- a (\partial_i \pi)^2\right) 
\end{equation*}
The Euler- Lagrange equation for the field $\pi$ becomes,
\begin{equation*}
\frac{\partial}{\partial t}\left( \frac{2 a^3 M_{pl}^2 \dot{H}}{c_s^2} \dot{\pi}\right) -2 M_{pl}^2 \dot{H}a \nabla^2 \pi = 0
\end{equation*}
Decomposing $ \pi $ into momentum modes using Fourier transform,
\begin{equation*}
\pi (\tau,\textbf{x}) = \int \mathrm{d^3}k \pi_{\textbf{k}}(\tau) e^{ i\textbf{k.x}}
\end{equation*}
we get
\begin{equation*}
\frac{\partial}{\partial t}\left( \frac{2 a^3 M_{pl}^2 \dot{H}}{c_s^2} \dot{\pi_{\textbf{k}}}\right) ~+~ 2 M_{pl}^2 \dot{H}a k^2 \pi_{\textbf{k}} = 0
\end{equation*}

Now defining the Mukhanov-Sasaki variable
\begin{equation*}
v_{\textbf{k}} = z \pi _{\textbf{k}}
\end{equation*}
where 
\begin{equation*}
z = \frac{a \sqrt{-2\dot{H}} ~ M_{pl}}{c_s}
\end{equation*}
and transforming everything to conformal time $ \tau $, such that $ dt ~=~ a ~d \tau$ we get
\begin{equation*}
v_{\textbf{k}}'' +\left(k^2 c_s^2 - \frac{z''}{z}\right)v_{\textbf{k}} = 0
\end{equation*}
where $'$ represents derivative with respect to conformal time. 
Now the strongest time dependence is contributed by scale factor a so  if we take $\dot{H}$ and $c_s$ to be varying  slowly, we can write $ \frac{z''}{z} = \frac{a''}{a} $. For perfect de Sitter space $ a = -\frac{1}{H \tau}~$ so $~ \frac{a''}{a} =  \frac{2}{\tau ^2} $.
\\
So the equation reduces to 
\begin{equation}
v_{\textbf{k}}'' +\left(k^2 c_s^2 - \frac{2}{\tau^2}\right)v_{\textbf{k}} = 0
\end{equation}

which has a solution
\begin{equation*}
v_{\textbf{k}} = \frac{-i(1 + i k c_s \tau)}{\sqrt{2 k c_s} ~c_s k \tau} e^{-ikc_s \tau}
\end{equation*}
For perturbation $\pi$, 
\begin{equation} \label{eq: pi}
\pi_{\textbf{k}} = \frac{v_k}{z} = \frac{i(1 + i k c_s \tau)}{2 \sqrt{\epsilon c_s k}~ k~M_{pl} } e^{-ikc_s \tau}
\end{equation}
where $\epsilon = -\frac{\dot{H}}{H^2} $ is the usual slow roll parameter and we have used $ a = -\frac{1}{H \tau}$. Differentiating it with respect to $\tau$ we get
\begin{equation} \label{eq: pi'}
\pi_{\textbf{k}}' = \frac{i}{2\sqrt{\epsilon k c_s} ~M_{pl}} (k c_s^2 \tau e^{-ikc_s \tau} )
\end{equation}
\\
Now we need to think of cubic terms in the action as the perturbation to quadratic action. In other words cubic terms will contribute to interaction part of the Hamiltonian. Upto the cubic order we have $ {\cal H}_{int} ~= ~ - {\cal L}_{int} $ \cite{Eugene} so the interaction Hamiltonian in leading order becomes

\begin{equation}
H_{int}(t) =  - \int \mathrm{d^3} x ~ a^3 \left[ M_{pl}^2 \dot{H}\left(1-\frac{1}{c_s^2}\right) \left( \dot{\pi}^3 - \dot{\pi}\frac{(\partial_i \pi)^2}{a^2}\right) -\frac{4}{3}M_3^4\dot{\pi}^3\right]
\end{equation}
This has two terms: one is $ a^3 \dot{\pi}^3 $ and the other one is $ a \dot{\pi}(\partial_i \pi)^2 $ . We will calculate the correlation function corresponding to each separately. 
\paragraph{Calculation for $ a^3 \dot{\pi}^3 $ term}
We can expand $\pi$ into creation and annihilation operators 
as
\begin{equation*}
\pi = \pi_{\textbf{k}} a_{\textbf{k}} + \pi_{\textbf{k}}^* a_{\textbf{k}} ^\dagger
\end{equation*}
Weinberg's in-in formalism \cite{Weinberg} gives us a very nice way to calculate correlation functions. According to it any correlation function $( \langle W(t) \rangle )$ is given by 
\begin{equation*}
\langle W(t) \rangle = \sum_{N~=~0}^{\infty} i^N 
\int_{t_0 }^t dt_N \int_{t_0}^{t_N} dt_{N-1} ... 
\int_{t_0}^{t_2} dt_1 \langle [H_{int} ( t_1),[H_{int}
(t_2),...[H_{int} (t_N) , W(t)]...]]\rangle . 
\end{equation*}
If we consider only upto first order in interaction we get
\begin{equation*}
\langle W(t) \rangle = i \int_{ t_0 }^t dt \langle [H_{int} (t) , W(t)] \rangle 
\end{equation*}

We will take $t_0$ to be the beginning of the universe which will become $-\infty$ once we convert everything to conformal time. We do the integral upto $\tau = 0$.
\begin{eqnarray}
\langle \pi({\textbf{k}}_1,0) \pi({\textbf{k}}_2,0) \pi({\textbf{k}}_3,0) \rangle =  -i \int_{-\infty}^{ 0} a \mathrm{d}\tau \int d^3 x  \langle  [ \pi({\textbf{k}}_1,0)  \pi({\textbf{k}}_2,0)
 \pi({\textbf{k}}_3,0) ,\pi '^3 (\tau,x) ] \rangle ~~~~
\end{eqnarray}
Now again decomposing $\pi'$ into Fourier modes

\begin{equation*}
\langle \pi({\textbf{k}}_1,0) \pi({\textbf{k}}_2,0) \pi({\textbf{k}}_3,0) \rangle =  -i(2\pi) ^3 \int_{-\infty}^{ 0} a \mathrm{d}\tau  \int d {\textbf{k}}_4 \int d {\textbf{k}}_5 \int d {\textbf{k}}_6  \langle  [ \pi({\textbf{k}}_1,0)  \pi({\textbf{k}}_2,0)
 \pi({\textbf{k}}_3,0) ,
\end{equation*}
\begin{equation*}
                                                 \pi'({\textbf{k}}_4,\tau)  \pi'({\textbf{k}}_5,\tau)  \pi'({\textbf{k}}_6,\tau) ] \rangle  \delta^3(\mathbf{k_4} +\mathbf{k_5} + \mathbf{k_6})
\end{equation*}

Now using Wick contraction and considering only connected diagrams
\begin{equation*}
\langle \pi({\textbf{k}}_1,0) \pi({\textbf{k}}_2,0) \pi({\textbf{k}}_3,0) \rangle = -6 i (2\pi)^3 \delta^3(\mathbf{k_1} +\mathbf{k_2} + \mathbf{k_3}) \int_{-\infty}^{0}  a \mathrm{d}\tau (\pi_{{\textbf{k}}_1}(0)) (\pi'_{{\textbf{k}}_1}(\tau)) ^{*} 
\end{equation*}
\begin{equation*}
\pi_{{\textbf{k}}_2}(0)) (\pi'_{{\textbf{k}}_2}(\tau)) ^{*} (\pi_{{\textbf{k}}_3}(0)) (\pi'_{{\textbf{k}}_3}(\tau)) ^{* } +   C.C.
\end{equation*}

Putting values of $ \pi_{\textbf{k}} $ and $ \pi_{\textbf{k}}' $ from Eq. \ref{eq: pi} and \ref{eq: pi'}, we get
\begin{equation*}
\begin{split}
\langle \pi({\textbf{k}}_1,0) \pi({\textbf{k}}_2,0) \pi({\textbf{k}}_3,0)\rangle ~=~ -\frac{3 i c_s^3}{32 
\epsilon^3 k_1 k_2 k_3 M_{pl}^6}(2\pi)^3 \delta^3(\mathbf{k_1} 
+\mathbf{k_2} + \mathbf{k_3}) \\ \int_{-\infty}^{0} a \tau^3 
\left[ e^{i (k_1 + k_2 + k_3) c_s \tau} - e^{- i(k_1 + k_2 + 
k_3) c_s \tau}\right]  \mathrm{d} \tau
\end{split}
\end{equation*}
or putting $ a = -\frac{1}{H \tau}$ we get

\begin{equation*}
\begin{split}
\langle \pi({\textbf{k}}_1,0) \pi({\textbf{k}}_2,0) \pi({\textbf{k}}_3,0)\rangle ~=~ +\frac{3 i c_s^3 }{32 \epsilon^3 H  k_1 k_2 k_3 M_{pl}^6}(2\pi)^3 \delta^3(\mathbf{k_1} +\mathbf{k_2} + \mathbf{k_3}) \\  \left[ \int_{-\infty}^{0}  \tau^2 e^{i (k_1 + k_2 + k_3) c_s \tau} \mathrm{d} \tau  - \int_{-\infty}^{0}  \tau^2 e^{i (k_1 + k_2 + k_3) c_s \tau} \mathrm{d} \tau \right] 
\end{split}
\end{equation*}

Now this integral doesn't converge. But we can get rid of divergences using a slightly imaginary time axis while doing integration as follows:
  \begin{equation*}
  \begin{split}
\langle \pi({\textbf{k}}_1,0) \pi({\textbf{k}}_2,0) \pi({\textbf{k}}_3,0)\rangle  = ~+~\frac{3 i c_s^3 }{32 \epsilon^3 H  k_1 k_2 k_3 M_{pl}^6}(2\pi)^3 \delta^3(\mathbf{k_1} +\mathbf{k_2} + \mathbf{k_3}) \\ \left[ \int_{-\infty(1-i\epsilon)}^{0}  \tau^2 e^{i (k_1 + k_2 + k_3) c_s \tau} \mathrm{d} \tau  - \int_{-\infty(1 + i \epsilon)}^{0}  \tau^2 e^{-i (k_1 + k_2 + k_3) c_s \tau} \mathrm{d} \tau \right]  
\end{split}
\end{equation*}
which gives
\begin{equation}
\langle \pi({\textbf{k}}_1,0) \pi({\textbf{k}}_2,0) \pi({\textbf{k}}_3,0)\rangle = ~-~\frac{3}{8 \epsilon^3 H k_1 k_2 k_3 K^3 M_{pl}^6}(2\pi)^3 \delta^3(\mathbf{k_1} +\mathbf{k_2} + \mathbf{k_3}) ~~~~
\end{equation}
where $K= k_1 + k_2 + k_3 $.

\paragraph{Calculation for $ a \dot{\pi}(\partial_i \pi)^2 $ term}
Now doing this calculation for $ a \dot{\pi}(\partial_i \pi)^2 $ term, we get 
\begin{eqnarray}
\langle \pi({\textbf{k}}_1,0) \pi({\textbf{k}}_2,0) \pi({\textbf{k}}_3 ,0) \rangle =  -i \int_{-\infty}^{ 0} a \mathrm{d}\tau \int d^3 x  \langle  [ \pi({\textbf{k}}_1,0)  \pi({\textbf{k}}_2,0)
 \pi({\textbf{k}}_3,0) ,\pi'(\partial_i \pi)^2 (\tau,x) ] \rangle
\end{eqnarray}

\begin{equation*}
\langle\pi({\textbf{k}}_1,0) \pi({\textbf{k}}_2,0) \pi({\textbf{k}}_3,0) \rangle =  -i (2 \pi)^3 \int_{-\infty}^{ 0} a \mathrm{d}\tau  \int d {\textbf{k}}_4 \int d {\textbf{k}}_5 \int d {\textbf{k}}_6  \langle  [ \pi({\textbf{k}}_1,0)  \pi({\textbf{k}}_2,0)
 \pi({\textbf{k}}_3,0) ,
\end{equation*}
\begin{equation*}
                      \pi'({\textbf{k}}_4,\tau)  \partial_i \pi ({\textbf{k}}_5,\tau)  \partial_i \pi ({\textbf{k}}_6,\tau) ]  \rangle  \ \delta^3(\mathbf{k_4} +\mathbf{k_5} + \mathbf{k_6})
\end{equation*}

Now using Wick contraction and considering only connected diagrams
\begin{equation*}
\begin{split}
\langle \pi({\textbf{k}}_1,0) \pi({\textbf{k}}_2,0) \pi({\textbf{k}}_3,0) \rangle =  i (2\pi)^3 \delta^3(\mathbf{k_1} +\mathbf{k_2} + \mathbf{k_3}) \int_{-\infty}^{0}  a \mathrm{d}\tau \\
[ 2 {\textbf{k}}_2 . {\textbf{k}}_3 (\pi_{{\textbf{k}}_1}(0)) (\pi'_{{\textbf{k}}_1}(\tau)) ^{*}  \pi_{{\textbf{k}}_2}(0)) (\pi_{{\textbf{k}}_2}(\tau)) ^{*} (\pi_{{\textbf{k}}_3}(0)) (\pi_{{\textbf{k}}_3}(\tau)) ^{* } \\
                   + 2 {\textbf{k}}_1. {\textbf{k}}_3 (\pi_{{\textbf{k}}_1}(0)) (\pi_{{\textbf{k}}_1}(\tau)) ^{*} \pi_{{\textbf{k}}_2}(0)) (\pi'_{{\textbf{k}}_2}(\tau)) ^{*} (\pi_{{\textbf{k}}_3}(0)) (\pi_{{\textbf{k}}_3}(\tau)) ^{* } \\
                   + 2 {\textbf{k}}_1.{\textbf{k}}_2 (\pi_{{\textbf{k}}_1}(0)) (\pi_{{\textbf{k}}_1}(\tau)) ^{*} \pi_{{\textbf{k}}_2}(0)) (\pi_{{\textbf{k}}_2}(\tau)) ^{*} (\pi_{{\textbf{k}}_3}(0)) (\pi'_{{\textbf{k}}_3}(\tau)) ^{* }] \\
                   + C.C.
\end{split}
\end{equation*}
Now since $\mathbf{k_1},\mathbf{k_2}$ and $\mathbf{k_3}$ form 
the sides of a triangle, $2 {\textbf{k}}_2 . {\textbf{k}}_3 = 
(k_1^2 ~-~ k_2^2 - k_3^2)$ and 
similarly for other two permutations. Now we put the values of 
$ \pi_{\textbf{k}} $ and $ \pi_{\textbf{k}}' $ from Eq. 
\ref{eq: pi} and \ref{eq: pi'}, and integrate. After some algebra we get, 
\begin{eqnarray}
\langle \pi({\textbf{k}}_1,0) \pi({\textbf{k}}_2,0) \pi({\textbf{k}}_3,0) \rangle = 
 -\frac{1}{32 \epsilon^3 c_s^2 H (M)^3  M_{pl}^6}(2\pi)^3 \delta^3(\mathbf{k_1} +\mathbf{k_2} + \mathbf{k_3})\nonumber
  \\ \left( \frac{ K^6 + 12 M^2 -4 ~K ~L ~M -4~K^2~L^2 + 11 M~ K^3 -3 L K^4 }{ K ^3 }\right) 
\end{eqnarray}
where 
\begin{eqnarray}
K ~=~ k_1~+~k_2~+~k_3 \\
L ~=~ k_1 k_2 ~+~ k_2 k_3 ~+~ k_3 k_1 \\
M ~=~ k_1 k_2 k_3  
\end{eqnarray}

\paragraph{Total Bispectrum:}
We can get total bispectrum by adding the above two terms multiplied by the coefficients in the original action.This gives:

\begin{equation*}
\begin{split}
\langle \pi(k_1,0) \pi(k_2,0) \pi(k_3,0)\rangle = -(2\pi)^3 \delta^3(\mathbf{k_1} +\mathbf{k_2} + \mathbf{k_3}) [ - \left( M_{pl}^2 \dot{H}\left(1-\frac{1}{c_s^2}\right) -\frac{4}{3}M_3^4\right)\frac{3}{8 \epsilon^3 H k_1 k_2 k_3 K^3 M_{pl}^6}   \\ 
~+~M_{pl}^2 \dot{H}\left(1-\frac{1}{c_s^2}\right)  \frac{1}{32 \epsilon^3 c_s^2 H (M)^3  M_{pl}^6}  \left( \frac{ K^6 + 12 M^2 -4 ~K ~L ~M -4~K^2~L^2 + 11 M~ K^3 -3 L K^4 }{ K ^3 }\right) ]   
\end{split}
\end{equation*}
In the effective field theory model this $\pi$ can be related to $\tau$ that remains unchanged after horizon crossing as
\begin{equation*}
\zeta (t,\textbf{x}) = -H \pi(t, \textbf{x})
\end{equation*} 
So,
\begin{equation*}
\langle \zeta(k_1,0) \zeta(k_2,0) \zeta(k_3,0)\rangle = -H^3 \langle \pi(k_1,0) \pi(k_2,0) \pi(k_3,0)\rangle
\end{equation*}
or
\begin{equation*}
\langle \zeta(k_1,0) \zeta(k_2,0) \zeta(k_3,0)\rangle = (2\pi)^3 \delta^3(\mathbf{k_1} +\mathbf{k_2} + \mathbf{k_3})\left[ F_{\dot{\pi}^3} (k_1,k_2,k_3) + F_{\dot{\pi}(\partial_i \pi)^2} (k_1,k_2,k_3)\right] 
\end{equation*}
where
\begin{equation}
F_{\dot{\pi}^3} (k_1,k_2,k_3) = - \left( M_{pl}^2 \dot{H}\left(1-\frac{1}{c_s^2}\right) -\frac{4}{3}M_3^4\right)\frac{3 H^2}{8 \epsilon^3  k_1 k_2 k_3 K^3 M_{pl}^6}
\end{equation}
and
\begin{eqnarray}
F_{\dot{\pi}(\partial_i \pi)^2} (k_1,k_2,k_3) = ~ \left(1-\frac{1}{c_s^2}\right) \times \frac{H^2 \dot{H} }{32 \epsilon^3 c_s^2  (M)^3  M_{pl}^4} \nonumber \\  \left( \frac{ K^6 + 12 M^2 -4 ~K ~L ~M -4~K^2~L^2 + 11 M~ K^3 -3 L K^4 }{ K ^3 }\right)
\end{eqnarray}

\paragraph{Power Spectrum :}  Let's now define power spectrum $ \Delta _{\zeta} $ as
\begin{equation*}
\langle \zeta(\mathbf{k_1},0) \zeta(\mathbf{k_2},0) \rangle = (2\pi)^3 \delta^3(\mathbf{k_1} -\mathbf{k_2})\frac{\Delta _{\zeta}(k)}{k_1^3}
\end{equation*}
Also 
\begin{equation*}
\langle \zeta(\mathbf{k_1},0) \zeta(\mathbf{k_2},0) \rangle = H^2 (2\pi)^3 \delta^3(\mathbf{k_1} -\mathbf{k_2})|\pi_{k_1}|^2|
\end{equation*}
or 
\begin{equation*}
\langle \zeta(\mathbf{k_1},0) \zeta(\mathbf{k_2},0) \rangle = H^2 (2\pi)^3 \delta^3(\mathbf{k_1} -\mathbf{k_2}) \frac{1}{4 \epsilon k^3 c_s M_{pl}^2}
\end{equation*}
So
\begin{equation*}
\Delta _{\zeta}(k) = \frac{H^2}{4 \epsilon k^3 c_s M_{pl}^2} 
\end{equation*}

\paragraph{Experimental constraints :}
We define $ f_{NL} $ as
\begin{equation*}
f_{NL} =\frac{5}{18} \frac{F(k,k,k)}{\Delta _{\zeta}(k)^2}
\end{equation*}
For $\dot{\pi}(\partial_i \pi)^2$ term 
\begin{equation*}
f_{NL}^{\dot{\pi}(\partial_i \pi)^2} = \frac{85}{324} \left(1-\frac{1}{c_s^2}\right) 
\end{equation*} 
and for $ \dot{\pi}^3 $ term
\begin{equation*}
f_{NL}^{\dot{\pi}^3} = \frac{15}{243} \left( c_s^2 - 1-\frac{4}{3}\frac{c_s^2 M_3^4}{M_{pl}^2 \dot{H}}   \right)  
\end{equation*}
Now this shape doesn't match equilateral or orthogonal shape. It is a combination of both.  The exact relation between $ f_{NL}^{\dot{\pi}(\partial_i \pi)^2}, f_{NL}^{\dot{\pi}^3} $ and $f_{NL} ^{equil.} ,f_{NL}^{orthog.} $ have been given in \cite{Senatore2} as 
\[ \left( \begin{array}{c}
f_{NL} ^{equil.} \\
 f_{NL}^{orthog.} \end{array} \right) =\left(\begin{array}{cc}
 1.040 & 1.210\\
 -0.03951 & -0.1757 \end{array}\right)  \left( \begin{array}{c}
f_{NL} ^{\dot{\pi}(\partial_i \pi)^2} \\
 f_{NL}^{\dot{\pi}^3} \end{array} \right)
\] 
Inverting these relations, we get
\begin{equation*}
f_{NL} ^{\dot{\pi}(\partial_i \pi)^2} = 1.3022 f_{NL} ^{equil.} + 8.9682 f_{NL}^{orthog.}  
\end{equation*}
\begin{equation*}
f_{NL} ^{\dot{\pi}^3} = -0.2928 f_{NL} ^{equil.} -7.7082 f_{NL}^{orthog.}
\end{equation*}
Planck \cite{Planck} sets the constraints as $-86 < f_{NL} ^{equil.} < 54 $ and $ -67 < 
f_{NL}^{orthog.} < -1 $ ( 68 \%  CL statistical) which implies constraints as

\begin{equation}
-712 < f_{NL} ^{\dot{\pi}(\partial_i \pi)^2} < 61
\end{equation}

\begin{equation}
-8.103 < f_{NL} ^{ \dot{\pi}^3 } < 541
\end{equation}

This gives for speed of sound
\begin{equation}
c_s > 0.019 
\end{equation}

\section{Conclusion}
We use the effective field theory model of inflation and analytically calculate the 2-
point and the 3-point functions for the field perturbations. Then we use the latest Planck 
data \cite{Planck} to constrain various parameters of the model. 

This similar calculation can be found in \cite{Senatore1} where authors have used WMAP data to 
constrain the parameters. Our final form of three point function matches with that. Moreover here we show all the calculation steps explicitly and use the latest Planck data. Also the constraint on speed of sound $ c_s$ was found by Planck collaboration in \cite{Planck} as $ c_s > 0.021$ for DBI model.

\section{Acknowledgements}
Authors thank the Indian Academy of Sciences for providing the opportunity to work as summer fellows and the ICTS-TIFR, 
Bangalore where most of this work was done. We also thank Dr. Suvrat Raju for guiding and 
supporting us through this project. We would also like to thank S.N. Bhatt fellowship 
program by ICTS which supported us.  We are also supported by INSPIRE fellowship from DST, 
Government of India.

\end{document}